\documentstyle[12pt]{article}
\newcommand{\bpsi}{\mbox{\boldmath $\psi$}}
\newcommand{\bphi}{\mbox{\boldmath $\varphi$}}

\newcommand{\btau}{\mbox{\boldmath $\tau$}}

\newcommand{\bldeta}{\mbox{\boldmath $\eta$}}

\newcommand{\ba}{\begin{eqnarray}}
\newcommand{\ea}{\end{eqnarray}}
\newcommand{\be}{\begin{equation}}
\newcommand{\ee}{\end{equation}}

\begin{document}
\begin{titlepage}
\begin{center}
\hfill TAN-FNT-95-08
\\
\hfill SI-95-TP3S2
\vspace*{2cm}

{\large \bf A $J^\pi = 0^-$ resonance for $\pi NN \rightarrow \pi NN $
in the Skyrme model}

\vskip 1.5 cm

{Bernd Schwesinger$^a$ and Norberto N. Scoccola$^{b,c}$
\footnote[2]{Fellow of the CONICET,
Buenos Aires, Argentina.}}
\vskip .2cm

{\it $^a$ Siegen University, Fachbereich Physik, 57068 Siegen,
Germany\\
$^b$ Departamento de F\'{\i}sica, CNEA, Av.\ Libertador 8250,
1429 Buenos Aires, Argentina.\\
$^c$ INFN, Sez.\ Milano, Via Celoria 16, 20133 Milano, Italy.}

\vskip 2.cm

July, 1995

\vspace{1.cm}

{\bf ABSTRACT}\\
\begin{quotation}
The scattering of pions on a dibaryon configuration is analyzed within
the $SU(2)$ Skyrme model. It is shown that this model leads to a low-lying
$(J^P,I)= (0^-,2)$ resonance. The possibility that this resonance
corresponds to one proposed recently in the context of double charge
exchange pion scattering on nuclei is discussed.
Given the setup used in those experiments we also show that
a resonance with isospin assignment $I=0$ cannot be excited according
to the description presented here.
\end{quotation}
\end{center}
\end{titlepage}

\noindent
A recent reevaluation of $(\pi^+,\pi^-)$ double charge exchange
cross sections over a range of nuclei has led to experimental evidence
for a $J^P=0^-$ resonance of the two nucleon system only 50 MeV above pion
threshold, i.e. far below the delta resonance\cite{BCS93}.
Its width has been estimated to amount to a few MeV only.
At present, it appears that this reaction produces the only clear candidate for
a dibaryon resonance standing out from a multitude of unclear signatures
in various channels. Although the arguments given in Refs.\cite{BCS93,BCS94}
seems to favor the assignment $I=0$ for the isospin of the resonance,
the case $I=2$ is not completely excluded\cite{GM93}.

Theoretical considerations of this specific resonance have been performed in
terms of the non-relativistic quark model assuming its structure could
be attributed to p-wave excitations of six valence quarks\cite{Faess94}.
Not surprisingly, the conclusion there is essentially negative, since
its description is the analogue of odd parity excitations in
non-relativistic three quark systems located around 600 MeV excitation
energy in the latter.

The Skyrme model, finally, is known to exhibit a non-trivial winding
number two configuration, the torus\cite{KS87,V87},
which has had, until recently, a
rather unclear interpretation ranging from ``an artifact of the model"
to ``the origin of intermediate nucleon-nucleon attraction", for the
most ambitious effort in latter direction see refs.\cite{Man95,Wal95}.
A recent investigation of the baryon number two configurations in large
$N_C$ chiral perturbation theory\cite{S95} has, however, shown that the
meson cloud around two explicit baryon sources indeed will assume the
shape of a torus, thus adding some weight to such toroidal configurations.

It is the purpose of the present letter, to examine the scattering of
pions from a torus-like meson cloud much in the spirit of pion-soliton
scattering\cite{WE84,HEHW84}, which has shown remarkable success for
cases where $1/N_C$
corrections are small: in those cases the leading $N_C$ contribution
suffices and explains gross features of physical pion-nucleon scattering
in the real world at $N_C=3$. It is our conjecture, that the leading
$N_C$ terms in the winding number two sector, which firstly lead to a
toroidal configuration at small internucleon distances\cite{S95},
will also be
able to capture the essence of this postulated resonance.

\noindent

The chiral fields of the axially symmetric $B=2$ configuration of
lowest energy, the torus, follow from a specific ansatz
\cite{WSH86,KS87,V87}
\begin{eqnarray}
U_{T}(\bf x )   & = & {\rm exp} \{ i \btau \cdot {\bf n} \, \chi \}
\end{eqnarray}
where the chiral angle $\chi =  \chi (r,\theta ) $ satisfies the
boundary conditions
\begin{eqnarray}
 \chi(r=0,\theta)= \pi, \qquad
\chi(r\rightarrow \infty, \theta) = 0,
\end{eqnarray}
and the orientation ${\bf n}({\bf x})$ of the pion fields
\begin{eqnarray}
{\bf n} & = & \left( \begin{array}{c}
\cos 2\varphi \, \sin \alpha (r,\theta )\\
\sin 2\varphi\, \sin \alpha (r,\theta )\\
\cos \alpha(r,\theta ) \end{array} \right)
\end{eqnarray}
rotates azimuthally twice as fast as for hedgehog configurations.
The symmetries of the torus are given by
\begin{eqnarray}
\chi(r,\theta) = \chi(r, \pi - \theta), \\
\alpha(r,\theta) = \pi - \alpha(r,\pi - \theta) \nonumber .
\end{eqnarray}
Chiral fields with respect to the physical isospin axes are obtained
after an isospin rotation $A \in {\rm SU}(2)$
\begin{eqnarray}
\btau \cdot {\bf n}({\bf x })\, \chi({\bf x})
\rightarrow
A \btau A^\dagger \cdot {\bf n}(R^{-1}\cdot {\bf r }) \,
\chi(R^{-1}\cdot {\bf r})
\end{eqnarray}
and the spatial coordinates ${\bf r}$ in the laboratory system are
related to the body-fixed coordinates ${\bf x}$ by a rotation $R$.
The construction of the rotational states of the torus\cite{WSH86} follows the
collective quantization procedure $A=A(t), R=R(t)$ and leads to a tower
of states
\begin{eqnarray}\label{rot}
\langle A,R \mid T\,T_3, \Sigma \,S_3, N_3\rangle =
(-)^{T-T_3} D^T_{-T_3 \, N_3}(A) D^\Sigma_{2N_3 \, S_3}(R)
\end{eqnarray}
where the symmetries $S_3^{bf} = 2 T_3^{bf}$ of the torus require that
body-fixed spin be twice as large as body-fixed isospin. Under a parity
transformation the chiral fields in the exponent transform according to
\begin{eqnarray}
{\bf P}\bigl( {\bf n}({\bf x })\, \chi({\bf x})\bigr) {\bf P}^{-1}=
-{\bf n}(-{\bf x })\, \chi(-{\bf x})=
\left(
\begin{array}{ccc}
-1 & & \\
 & -1 & \\
 & & 1
\end{array}
\right) \cdot
{\bf n}({\bf x })\, \chi({\bf x})
\end{eqnarray}
thus being equivalent to an isospin-rotation around the 3-axis by an
angle of $\pi$. This leads to the conclusion that the parity of a given
rotator state is carried by the projection of body-fixed isospin on the
symmetry axis of the torus
\begin{eqnarray}
P_{rot}= (-)^{N_3}\quad .
\end{eqnarray}
{}From the spins, isospins and parities of the rotational excitations we
conclude, that there is no state of negative parity and spin zero
present in purely rotational excitations of the torus. These additional
degrees of freedom can only come from intrinsic vibrational excitations
to which we turn now.

\noindent
Small amplitude oscillations of the torus are conveniently parametrized
by
\begin{eqnarray}\label{utexc}
U_{T^*}(\bf x )   & = & {\rm exp} \{ i \btau \cdot ( {\bf n} \, \chi \
\, + \, \bldeta ) \}
\end{eqnarray}
where the fluctuations can always be written in terms of a scaling
ansatz \cite{HH84}
\begin{eqnarray}\label{fluc}
\btau \cdot \bldeta  = \sum_M (-)^M \tau_{-M}
\Bigl( {\bf s}( {\bf x},t ) \cdot \nabla \Bigr)_{K_3}^P \, n_M( \hat x )
\chi ({\bf x})
\end{eqnarray}
for the fluctuating fields. A complete set of displacement fields ${\bf s}$
can be constructed from a set of three basis vectors
\begin{eqnarray}
\label{disp}
\left\{ {\bf s}({\bf x} )\right\} =
\left\{ {\bf x } , \nabla ,{\bf x } \times \nabla \right\} \sum_K
Y^K_{K_3}( \hat x ) F_K(r) \, .
\end{eqnarray}
multiplied by a harmonic time dependence.

\noindent
The additional parity $P$ carried by the vibration is determined by the parity
of the scaling operator:
\begin{eqnarray}
{\bf P }\left({\bf s}({\bf x} ) \cdot \nabla \right)^P_{K_3} {\bf P }^{-1} =
-\left({\bf s}(-{\bf x} ) \cdot \nabla \right)^P_{K_3} =
P \left({\bf s}({\bf x} ) \cdot \nabla \right)^P_{K_3}.
\end{eqnarray}
\noindent
Since the torus has no spherical symmetry the displacement fields of
different multipolarity $K$ do not decouple, however, their projection
$K_3$ on the symmetry-axis does. Arbitrary displacement fields can
therefore be expanded into an infinite sum of multipoles $L$
\begin{eqnarray}\label{multip1}
\bigl( {\bf s}({\bf x} ) \cdot \nabla \bigr)_{K_3}^P \, n_M( \hat {\bf x} )
\chi ({\bf x})
= \sum_L F^L_{K_3 \, P \, M}(r) \, Y^L_{K_3+2M}(\hat x ) \, .
\end{eqnarray}

\noindent
Generally, soliton fluctuations
will be unbound, i.e. above pion threshold, in which case the
displacements fields must grow exponentially with $m_\pi r$ in order to
compensate for the exponential decrease of the chiral angle $\chi$.
However, for an estimate of the resonance energies,
a polynomial form for the displacement fields
that keeps the vibrations localized in the vicinity of the torus
can be used\cite{HH84}. In our numerical calculations below
we will use such an approximation.

Since the ${\bf x}$ in e.g. eq.(\ref{multip1}) are the
components of coordinates with respect to the
body-fixed symmetry axis of the torus they do not contribute to the spin
of the configuration and since the isospin matrices in
eqs.(\ref{utexc},\ref{fluc}) also refer
to the body-fixed axes they do not contribute to isospin. Total spin
$J \, J_3$ and total isospin $I \, I_3$ of the configuration of
chiral fields for the vibrating torus  must be carried by the
Euler angles\cite{HEHW84} which indicate the
orientation of the laboratory coordinates ${\bf r } = R \cdot {\bf x}$
relative to the body-fixed ones
and of the isorotation $A$ necessary to
transform body-fixed matrices to the laboratory $ \btau \rightarrow
A \btau A^\dagger$. For a rigidly rotating body the (unnormalized)
quantized states are given by a generalization of eq.(\ref{rot})
\begin{eqnarray}\label{target}
\langle A,R \mid I \, I_3 , J \, J_3 \rangle =
(-)^{I-I_3} D^I_{-I_3 \, N_3}(A) \, D^J_{2N_3+K_3 \, J_3}(R)
\end{eqnarray}
where the indices $N_3$ and $2 N_3+K_3$ are fixed by the
symmetry constraints of the rotating object as will be seen later.
The full expression for the body-fixed fluctuations is summarized by
\begin{equation}\label{scattint}
\btau \cdot \bpsi  = \sum_M (-)^M \tau_{-M}
\Bigl\{
\Bigl( {\bf s}( {\bf x},t ) \cdot \nabla \Bigr)_{K_3}^P \, n_M( \hat x )
\chi ({\bf x}) \Bigr\}
\Bigl\{
 (-)^{I_3} D^I_{-I_3 \, N_3}(A) \, D^J_{2N_3+K_3 \, J_3}(R)
\Bigr\} \, .
\end{equation}
Expressed in terms of space
coordinates ${\bf r}$ of the laboratory system the scaling vibration reads
\begin{eqnarray}\label{multipole}
\bigl( {\bf s}({\bf x} ) \cdot \nabla \bigr)_{K_3}^P \, n_M( \hat x )
\chi ({\bf x})
& = &\sum_L
F^L_{K_3 \, P \, M}(r) \, Y^L_{K_3+2M}(\hat x ) \\
& = & \sum_{L \,L_3}
F^L_{K_3 \, P \, M}(r) \,  (-)^{K_3+2M-L_3}
D^L_{-2M-K_3 \quad \! -L_3}(R)
\, Y^L_{L_3}(\hat r ) \nonumber \, .
\end{eqnarray}
We now rotate the body-fixed fluctuation in eq.(\ref{scattint})
to physical isospin
axes, i.e. to the laboratory system, also inserting eq.(\ref{multipole})
\begin{eqnarray}\label{scattlab}
\btau \! \! \! \!& \! \! \cdot \! \! & \! \! \! \! \bphi =
A \btau \cdot \bpsi A^\dagger\\ \quad \! & = & \!
\sum_{m\,M} \tau_{m} D^1_{m \,-M}(A) (-)^M
\Bigl\{
\sum_{L \,L_3}
F^L_{K_3 \, P \, M}(r) \,  (-)^{K_3+2M-L_3}
D^L_{-2M-K_3 \quad \! -L_3}(R) \, Y^L_{L_3}(\hat r )
\Bigr\} \nonumber \\
 & & \qquad \qquad \qquad \qquad \qquad \qquad \qquad
\times \Bigl\{
(-)^{I_3} D^I_{-I_3 \, N_3}(A) \, D^J_{2N_3+K_3 \, J_3}(R)
\Bigr\} \nonumber \, .
\end{eqnarray}
Generally the energies of the
vibrations are above pion threshold and thus in the continuum: in this
case we speak of pion-torus scattering at fixed total spin
and isospin. In order to make the different scattering channels more
explicit we can recouple the $D$-functions of same Euler
angles in eq.(\ref{scattlab}) to
a single $D$-function, which then will represent the two-baryon
target state  coupled with isospin one and angular momentum $L$ of the
scattered pion:
\begin{eqnarray}\label{channels}
\btau \cdot \bphi & = &
(-)^{K_3} \sum_{L \, \Sigma \, T \,M}(-)^{I-T+L-M} \frac{\hat \Sigma
\hat T}{\hat J \hat I } \\
& &
    \left( ^{1}_{M}  \, ^{I}_{N_3} | ^{~~~T}_{N_3+M} \right)
    \left( ^{~~~~~L}_{-K_3+2M} \, ^{~~~~J} _{2N_3+K_3} |
^{~~~~\Sigma}_{2N_3+2M} \right)
F^L_{K_3 \, P \, -M}(r) \nonumber \\
& &\left[ \btau \circ D^T_{\cdot \, N_3+M}(A) \right]_{I\,I_3}
\left[D^{\Sigma}_{2N_3+2M \, \cdot}(R) \circ
Y^L_{\cdot}(\hat r ) \right]_{J\,J_3}
 \nonumber .
\end{eqnarray}
In eq.(\ref{channels}) we can identify the different scattering channels
\begin{eqnarray}
\left[ \btau \circ D^T_{\cdot \, N_3+M}(A) \right]_{I\,I_3}
\left[D^{\Sigma}_{2N_3+2M \, \cdot}(R) \circ
Y^L_{\cdot}(\hat r ) \right]_{J\,J_3}
\end{eqnarray}
where pionic isospin one  and target isospin $T$ are coupled to
total $ I \,I_3$ and pionic
angular momentum $L$ with  target spin $\Sigma$ to total $ J \, J_3$.
All target states in these coupled channels obey the
symmetry constraints of the torus: the third component of
body-fixed spin must be twice as large as body-fixed isospin. The
constraint is satisfied because we
had made the correct choice for the indices of the Euler angle
wavefunctions at the beginning in eq.(\ref{target}).
When the torus is quantized
according to two identical fermions an additional constraint on the
target
\begin{eqnarray}
(-)^{\Sigma + T} = -1
\end{eqnarray}
emerges because of the Pauli exclusion principle.
{}From the full expression for the fluctuations around the torus we see that
there is no restriction on the multipolarity $L$ of the pion and
the target spins $\Sigma$,
\begin{eqnarray}
 \mid L-\Sigma \mid \leq J \leq L+\Sigma \quad ,
\end{eqnarray}
quite in contrast to the
$B=1$ Skyrmion fluctuations. There the possible isospins of the target
$ \mid T-1 \mid \leq I \leq T+1 $ limit the possible spins because
hedgehogs obey the stronger symmetry constraint $ T = \Sigma$.

\noindent
Since the right index $N_3$ of the rotator state in eq.(\ref{target}) carries
parity $(-)^{N_3}$ we can read off
\begin{eqnarray}
P_{{\rm total}}= (-)^{N_3}\,P \quad
\end{eqnarray}
as total parity which will label the scattering process considered.

\noindent
Having given the general expressions corresponding to the scattering
of pions on the torus we will now study the particular configurations
used in the double charge exchange experiments.
The experimental setup has singled out one specific incoming channel
where an $L=0$ pion is incident an a two-nucleon pair in a spin $\Sigma
= 0$ state. This spin was chosen by the fact that the two nucleons,
being close together, cannot have any
relative angular momentum and their isospin must be $T=1$ in double
charge exchange experiments. Since total spin $J=0$ was observed by the
angular distributions of the cross sections the incoming pionic angular
momentum is fixed.

\noindent
Checking the intrinsic fluctuations with respect to the possible pionic
angular momenta $L$ for fixed $K_3$, eq.(\ref{multipole}), we deduce, that
$L=0$ can only occur for\\
(i) $K_3=0$ in the third component $M=0$ of the body-fixed torus fields
${\bf n}$, or\\
(ii) $\mid K_3 \mid =2$ for the first two components of ${\bf n}$ with
$\mid M \mid = 1$.\\
These two possibilities exhaust the set of $\pi NN$ incoming
channels to which all others are coupled according to eq.(\ref{channels}).

\noindent
Consider case (i) now: since the incoming two-nucleon spin $\Sigma$ is
zero, the parity of the incoming rotational motion must be positive, because
$\mid N_3 \mid \leq \Sigma =0$.
The negative parity of the incoming channel must be
attributed to the vibrational excitation: so here the incoming channel
is based on a $K_3^P = 0^-$ vibration. From Table 1 we conclude,
however, that the multipole decomposition of the $K_3^P=0^-$
vibration begins only above $L\geq 2$. In fact, we see that given
the corresponding form of ${\bf s}({\bf x})$ we have
\be
\bigl( {\bf s}({\bf x} ) \cdot \nabla \bigr)_{K_3=0}^{P=-1}\, {\bf n}
( {\bf x } ) = z \ \partial_\varphi \
\left( \begin{array}{c} \cos 2 \varphi\ \sin \alpha (r,\theta) \\
                        \sin 2 \varphi\  \sin \alpha (r,\theta) \\
                         \cos \alpha (r,\theta) \end{array} \right)
\quad ,
\ee
the components of which can be expressed in terms of the $Y^2_{\pm 2}$
spherical harmonics since the third component (M=0) vanishes.
So case (i) does not couple to
the incoming channel required by the experimental setup.

\noindent
This leaves case (ii) as unique possibility. Since the intrinsic
vibration in the incoming $L=0$ channel must have $\mid M \mid =1$
we also must have $\mid N_3 \mid =1$ for an incoming target state of
$\Sigma =0$. Hence total negative parity stems from the incoming target
state and the intrinsic vibration must carry $\mid K_3\mid ^P=2^+$.
This intrinsic vibration therefore just corresponds to a mode, which
tends to separate the torus into two $B=1$ solitons and which turns out
to be rather low in energy. We have estimated its energy using the
scaling ansatz discussed earlier and the basis for the displacement
fields given in Table 1. We have multiplied each element of the basis
by an arbitrary polynomial in $r$ and $z^2$ searching for minimal
excitation energy. In this way we obtained the estimate $\omega = 250$ MeV,
which is $110$ MeV above pion threshold. In this estimate
we have used the standard Skyrme model parameters for scattering in the
$B=1$ sector: $f_\pi=93$ MeV, $m_\pi=138$ MeV, $e=4$.

\noindent
Even if one relaxes the initial constraint: incoming pions in an $L=0$
channel, one can show by analogue arguments that $L=1$ channels
cannot couple to $J^P=0^-$ reactions on two nucleons. Finally,
$L\geq 2$ channels require $\Sigma \geq 2$ target states, inaccessible
from two-nucleon states. Thus the low-lying separation mode of the
torus is the door-way channel via which a $J^P=0^-$ scattering process
on two nucleons must proceed.

\noindent
Till now we have left aside all considerations with respect to isospin,
to which we turn now. Since the incoming channel specified in case (ii)
must have $\mid N_3 \mid =1$ the Clebsch-Gordon coefficient in the
channel decomposition of the scattering states, eq.(\ref{channels}),
forces total isospin $I$ to be greater than zero leaving $I=1,2$ for
incoming channels containing two nucleons.
The case of $I=1$ has been ruled out in the interpretation of the
scattering experiment because it leads to a large decay width to
the channel $\pi NN \rightarrow NN $, leaving only $I=0,2$ as possibilities.
Thus, the Skyrme model scattering states of pions on two nucleons would
coincide with the experimental determination of scattering quantum
numbers only if total isospin $I$ is uniquely $I=2$.

\noindent
Having clarified the role of the separation mode of the torus in
$J^P=0^-$ scattering of a pion on two nucleons, we still have to
convince ourselves, that this mode will actually lead to a narrow
resonance roughly 50 MeV above pion threshold. In principle, one would
have to calculate the pion-torus scattering amplitudes the same way as
it had been done for pion-soliton scattering\cite{WE84,HEHW84},
the latter with
remarkable success for those channels, where higher order contributions
in $1/N_C$ are unimportant. Avoiding this formidable effort, at least
for the moment, we resort to estimates. One estimate presented here was
based on the eigenfrequencies of prescribed scaling modes, which for the
door-way vibration under consideration came out at 110 MeV above pion
threshold. Alternatively, one may consult the two-soliton calculations
for the deuteron based on the Atiyah-Manton ansatz\cite{AM89}, which find the
separation mode as softest mode - apart from the zero modes, of course
- at 130 MeV. However, latter calculation could only be performed for
massless pions, so the relation between the two estimates given is somewhat
unclear. As for the width we have no handle for the moment, apart from
actually doing the pion-torus scattering. But, if a resonance will emerge at
roughly the estimated position it must have a small width, because it
is a resonance close to threshold.
We conclude by emphasizing that within the approximations and version
of the $SU(2)$ Skyrme model used here,
the isospin assignment $I=0$ for a $J^P=0^-$ dibaryon resonance seems
to be excluded given the setup used in the experiments. A resonance
with isospin $I=2$, however, appears very naturally in form of a
separation mode of the underlying toroidal $B=2$ soliton configuration.

\vspace{2.cm}

We wish to thank Niels Walet for explaining to us the
structure of his lowest vibrational mode, i.e. the separation mode of
the torus. This work was initiated while we were
participating at the INT-95-1 session on ``Chiral Dynamics in Hadrons
and Nuclei" at the Institute for Nuclear Theory at the University of
Washington, USA. We wish to thank the organizers of the session
for the invitation to participate and the Department of Energy, USA
for partial financial support during that period.
One of us, B.S., wishes to thank the
Istituto Nazionale di Fisica Nucleare for its hospitality allowing us
to complete the investigation.

\pagebreak

\newpage

\vspace*{3cm.}
\begin{center}
{\Large \bf Table 1}
\vspace{1cm}

\begin{tabular}{|c|c|}
\hline
                          &                                        \\
 $\mid K_3 \mid ^P = 0^-$ &
$\left( \begin{array}{c} y \ z \\ - x \ z \\0 \end{array} \right)$ \\
                          &                                        \\
\hline
                          &                                        \\
$\mid K_3 \mid ^P = 2^+$  &
$\left( \begin{array}{c} x \\ - y \\0 \end{array} \right)
\! \quad ; \quad \!
(x^2-y^2) \left( \begin{array}{c} x \\ y \\0 \end{array} \right)
\! \quad ; \quad \!
 x y \left( \begin{array}{c} x \\ - y \\0 \end{array} \right)
\! \quad ; \quad \!
(x^2-y^2) \left( \begin{array}{c} 0 \\ 0 \\z \end{array} \right)$ \\
                          &                                        \\
\hline
\end{tabular}
\vspace{0.5cm}

\begin{quotation}%\newcommand{\vr}{\vec r}
 [Table 1:]
\noindent
Basic building blocks for the displacement fields ${\bf s}({\bf x})$ in the
$\mid K_3 \mid ^P = 0^-$ and $2^+$ channels, as explained in the text.
These building blocks are obtained using
eq.(\ref{disp}) for the lowest possible values of $K$, disregarding
all combinations which decouple from them.
\end{quotation}

\end{center}

\end{document}